\documentclass[a4paper]{article}
\usepackage{tipa}
\usepackage{hyperref}
\usepackage{INTERSPEECH2021}
\usepackage{tikz}

\title{DiCOVA Challenge: Dataset, task, and baseline system for COVID-19 diagnosis using acoustics}
\name{Ananya Muguli$^\dagger$, Lancelot Pinto$^\ddagger$, Nirmala R.$^\dagger$, Neeraj Sharma$^\dagger$, Prashant Krishnan$^\dagger$,\\\textit{Prasanta Kumar Ghosh$^\dagger$, Rohit Kumar$^\dagger$, Shrirama Bhat$^\mp$,Srikanth Raj Chetupalli$^\dagger$,}\\\textit{Sriram Ganapathy$^\dagger$, Shreyas Ramoji$^\dagger$, Viral Nanda$^\ddagger$}\thanks{Thanks to the Department of Science and Technology, Government of India.}}
\address{$^\dagger$Indian Institute of Science, Bangalore, $^\ddagger$P.~D.~Hinduja Hospital, Mumbai, $^\mp$KMC Hospital, Mangalore}
\email{sriramg@iisc.ac.in}

\ninept
\begin{document}

\maketitle
\begin{abstract}
The DiCOVA challenge aims at accelerating research in diagnosing COVID-19 using acoustics (DiCOVA), a topic at the intersection of speech and audio processing, respiratory health diagnosis, and machine learning. This challenge is an open call for researchers to analyze a dataset of sound recordings, collected from COVID-19 infected and non-COVID-19 individuals, for a two-class classification. These recordings were collected via crowdsourcing from multiple countries, through a website application. The challenge features two tracks, one focusing on cough sounds, and the other on using a collection of breath, sustained vowel phonation, and number counting speech recordings. In this paper, we introduce the challenge and provide a detailed description of the task, and present a baseline system for the task.
\end{abstract}

\section{Introduction}
The COVID-19 pandemic has emerged as a significant health crisis. At the time of writing ($15-$June-2021), more than $175$ million cases and more than $3.8$ million casualties have been reported by the World Health Organization (WHO) from about $200$ countries across the world \cite{who}. Physical distancing and implementation of wide-scale population testing have served as key measures to contain the pandemic. The testing methods in use can be broadly divided into molecular and antibody testing. In molecular testing, chemical reagents are used to detect the constituents, like nucleic acids and proteins, of the SARS-CoV-2 virus in an individuals' throat or nasal swab sample. The reverse transcription polymerase chain reaction (RT-PCR) is one such testing method, and currently serves as a gold standard for COVID-19 testing.
However, cost of machinery, time, and expertise have limited the scalability of this method. The rapid antigen test (RAT) is another molecular testing method which alleviates the time limitation of RT-PCR but has high false negatives (low specificity). The swab based tests and molecular tests also violate physical distancing between participant and the health worker, posing a serious practical challenge. In summary, there is a need to discover alternative methodologies to diagnose COVID-19 infection that are efficient in terms of time, cost, and ease, allowing scalability.

\begin{figure}
    \centering
    \includegraphics[width=7.5cm, height=4cm]{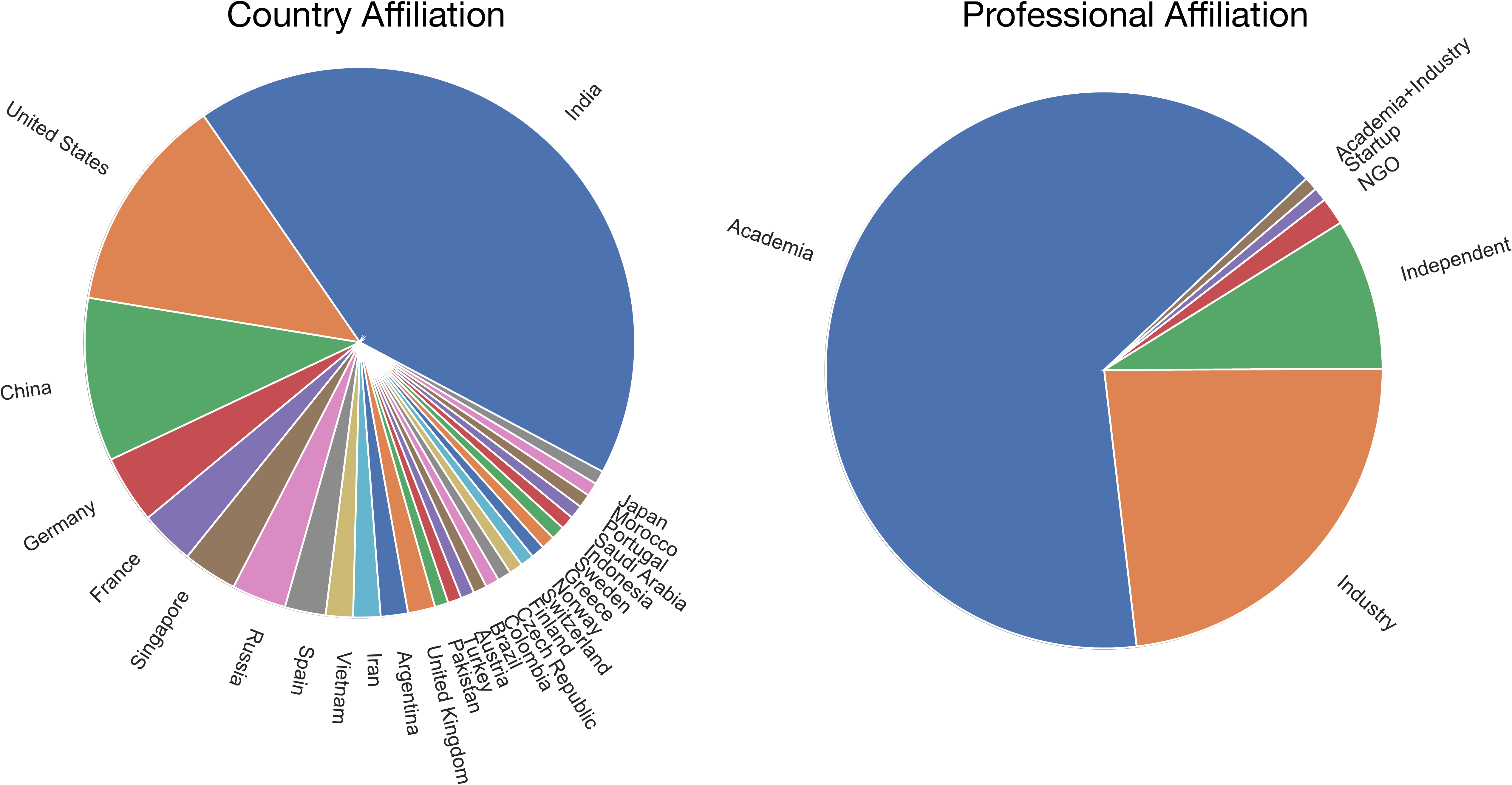}
    \caption{Illustration of the distribution $80$ plus challenge registrants (or teams).}
    \label{fig:chall_participation}
\end{figure}

\begin{figure*}
    \centering
    \includegraphics[width=17cm, height=9cm]{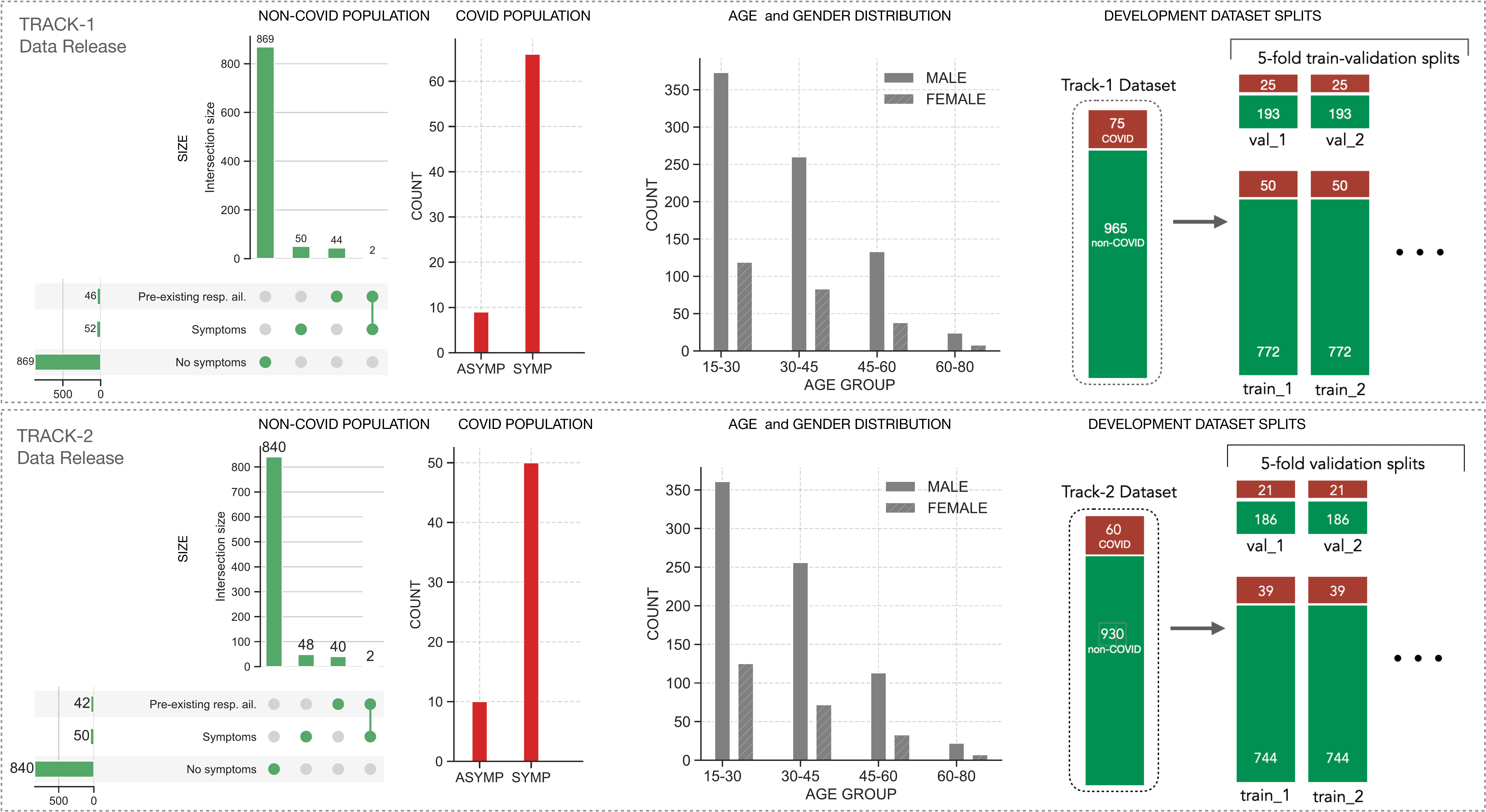} \caption{In each track, the dataset is grouped non-COVID and COVID subjects. The non-COVID subjects are either healthy, have symptoms (cough/cold), or have pre-existing respiratory ailments (chronic lung disease, asthma, or pneumonia). The COVID subjects are either symptomatic or asymptomatic COVID positive. The distribution of age, gender, and the splits of the development dataset is also shown.}
    \label{fig:baseline_tracks_metadata}
\end{figure*}

The WHO \cite{who} has maintained dry cough, breathing difficulty, chest pain, and fatigue as symptoms of the infection, manifested between $2-14$ days after exposure to the virus. This was also validated by a modeling study that analyzed data pertaining to the symptoms reported by $7178$ COVID-19 positive individuals \cite{smell_metadata}. The chest X-ray (and CT) scans of many COVID-19 infected individuals have revealed infection in the lungs \cite{islam20}, and effort is being directed to evaluate the feasibility of early diagnosis using imaging techniques. Interestingly, respiratory medical literature suggests that sounds emanating through coordinated release of air pressure through the lungs, such as breathing, cough, and speech, are intricately tied to changes in the anatomy and the physiology of the respiratory system \cite{speech_breathing_book}. A lung infection can affect the inspiratory and expiratory capacity. This, in addition to the presence of cough, can result in difficulty in vocalizing sustained phonation and/or continuous speech \cite{lee1993speech, chang2004perceived}. This has been the scientific principle based on which studies analyzing vocal sounds have shown some success in detecting respiratory ailments, such as pertussis \cite{pramono2016cough}, chronic obstructive pulmonary disease (COPD) \cite{windmon2018tussiswatch}, and tuberculosis \cite{botha2018detection}.

Based on such biological plausibility, we hypothesize that the evaluation of the accuracy of detecting COVID-19 using the acoustics of respiratory sounds merits research.
A success can provide an excellent point-of-care, quick, easy to use, and cost-effective tool to diagnose COVID-19 infection, and consequently contain COVID-19 spread. Altogether, it can supplement the molecular testing methods for COVID-19 detection or screening. The DiCOVA Challenge\footnote{\url{http://dicova2021.github.io/}} is designed to accelerate research efforts along this direction by creation and release of an acoustic signal dataset, and inviting researchers to build detection models and report performance on a blind test set.
\noindent Since its release on $04-$Feb-$2021$, the DiCOVA Challenge has created a widespread interest amongst researchers. We have received registration from more than $80$ teams. These come from various countries and professional affiliation (see Figure~\ref{fig:chall_participation}).
In this paper, we present an overview of the topic, tasks in the challenge, and the baseline system.

\section{Literature Review}
\label{literature}
Since the onset of the COVID-19 pandemic, several attempts are being made to evaluate the potential of sound based screening (and diagnosis). These attempts \cite{covid19sounddetector, coughagainstcovid20, breatheforscience20, coughvidepfl, covid19cmuproject, imran2020ai4covid, brown2020exploring, bagad2020cough} have primarily focused on cough sounds, and are work in progress. Brown et.al. \cite{brown2020exploring}
use cough and breathing sounds from $141$ COVID-19 patients, extract a collection of short-time frame-level acoustic features and embeddings from a VGGish network, and pass these through a logistic regression classifier. An area-under-the-curve (AUC) $~80\%$ is reported. The study by Imran et al.~\cite{imran2020ai4covid} uses sound samples from $48$ COVID-19 patients, and reports a sensitivity of $~94\%$ (and ~$91\%$ specificity) using a convolutional neural network (CNN) architecture, fed with mel-spectrogram features as the input. The study by Bagad et.al. \cite{bagad2020cough} uses cough samples from $376$ COVID-19 patients, and a CNN architecture based on ResNet18 with short-time magnitude spectrogram as input, and reports an AUC of $72\%$. Altogether, these studies are encouraging. The limitations include: $(i)$ a different COVID-19 patient population used in each study, $(iii)$ varied evaluation methodology, $(iii)$ small population size, and $(iii)$ lack of insight on acoustic feature differences between healthy and COVID-19 individuals. The DiCOVA Challenge is aimed to encourage multiple research groups to analyze the same dataset, evaluate the system performance using fixed metrics, and facilitate obtaining benchmarks for future system development.

\section{Dataset}
\label{dataset}
The DiCOVA Challenge dataset is derived from the Coswara dataset \cite{sharma2020coswara}, a crowd-sourced dataset of sound recordings from COVID-19 positive and non-COVID-19 individuals. The Coswara data is collected using a web-application\footnote{\url{https://coswara.iisc.ac.in/}}, launched in April-$2020$, accessible through the internet by anyone around the globe. The volunteering subjects are advised to record their respiratory sounds in a quiet environment. Each subject provides $9$ audio recordings, namely, $(a)$ shallow and deep breathing ($2~$nos.), $(b)$ shallow and heavy cough ($2$~nos.), $(c)$ sustained phonation of vowels [\ae] (as in bat), [i] (as in beet), and [u] (as in boot) ($3$~nos.), and $(d)$ fast and normal pace $1$ to $20$ number counting ($2$~nos.). The subjects also provided metadata corresponding to their current health status (includes COVID-19 status, any other respiratory ailments, and symptoms), demographic information like age and gender. From this Coswara dataset, we have created two datasets:
$(a)$ Track-1 dataset: composed of cough sound recordings,
and $(b)$ Track-2 dataset: composed of deep breathing, vowel [i], and number counting (normal pace) speech recordings.

\subsection{Metadata}
For the challenge, the subjects have been divided into two groups, namely, 
\begin{itemize}
    \item non-COVID: Subjects who are either healthy or have symptoms such as cold or cough, or have pre-existing respiratory ailments (asthma, pneumonia, chronic lung disease), and confirm that they are not COVID-19 positive.
    \item COVID: Subjects who confirm as COVID-19 positive, either symptomatic and asymptomatic.
\end{itemize} 
The Track-1 and Track-2 development datasets are composed of $1040$ ($965$ non-COVID subjects) and $990$ ($930$ non-COVID subjects), respectively. A breakdown of the subject population with respect to symptoms, age group, and gender is shown in Figure~\ref{fig:baseline_tracks_metadata}.

\subsection{Audio}
The Coswara data collection is via crowd-sourcing, which means the quality of the audio files has high variability and serves as a good representation of audio data collected in the wild. A majority of the audio files are clean as confirmed via informal listening.
More than $90\%$ of the collected files have a sampling rate of $48~$kHz and stored as WAV files. For the challenge datasets, all audio recordings have been re-sampled to $44.1$~kHz and compressed as FLAC format files. The Track-1 audio files correspond to cough sound signals. Each audio file is derived from one unique subject, and has one or more cough bouts. In total, there are $1040$ recordings. The average duration of recordings across subjects is $4.72 ($standard error $\pm 0.07)$~sec. The Track-2 audio files correspond to one of the three different sound categories, namely, breathing, vowel [i], and $1$ to $20$ number counting. In total, there are $3$(categories)$\times 990$ (subjects) sound recordings in Track-2. The average duration of recordings across subjects is: breath $17.72 (\pm 0.68)$~sec, vowel [i] $12.40 (\pm 0.17)$~sec, and number counting speech $14.71 (\pm 0.11)$~sec.

\section{Challenge Tasks}
\label{challenge}
The DiCOVA challenge features two tracks. Below we present the task and the instructions associated with each track. A participant can choose to participate in one or both the tracks.
\subsection{Track-1}
The goal is to use cough sound recordings from COVID-19 and non-COVID-19 individuals for the task of COVID-19 detection.
\begin{itemize}
\item The Track-1 development dataset is composed of cough audio data from $1040$ subjects. The dataset also contains lists corresponding to a $5-$fold cross validation split. The distribution of COVID and non-COVID in these splits is shown in Figure~\ref{fig:baseline_tracks_metadata}(a).
All participants are required to adhere to these lists and report the average performance over the $5$ validation sets.
\item A separate blind evaluation dataset is provided to all participants. The participants are required to report their COVID-19 detection scores as probabilities.
\item This is the primary track for the challenge. A baseline system is provided, and an online leaderboard \footnote{https://competitions.codalab.org/competitions/29640\#results} is set up for all participants to report and compare their performance.
\end{itemize}

\subsection{Track-2}
The goal is to use breathing, sustained phonation, and speech sound recordings from COVID-19 and non-COVID-19 individuals for any kind of detailed analysis which can contribute towards COVID-19 detection.
\begin{itemize}
\item The Track-2 development dataset is composed of three sets of sound recordings, namely, breathing, vowel [i], and number counting, from $990$ subjects.
\item The dataset also contains $5$ train-validation splits.
The distribution of COVID and non-COVID in these splits is shown in Figure~\ref{fig:baseline_tracks_metadata}(b).
\item The participants are encouraged to design COVID-19 detection systems using above splits.
\item This track has no baseline system and leaderboard. A non-blind test set is provided to all participants.
\end{itemize}

\noindent Participants are free to use any other data except the publicly available Project Coswara dataset \footnote{\url{https://github.com/iiscleap/Coswara-Data}}
for data augmentation, transfer learning, etc.

\subsection{Performance Evaluation}
Both Track-1 and Track-2 are binary classification tasks. With a focus on COVID detection, the performance is evaluated using the traditional detection metrics, namely, true positive (TP) and false positive (FP) rates, over a range of decision thresholds between $0-1$ with a step-size of $0.0001$. For track-1, the participant is required to submit a COVID probability score for every audio file (corresponding to a subject) in the blind test set. In the evaluation, we use the probability scores to compute the receiver operating characteristic (ROC) curve, and use the area under the curve (AUC) to quantify the model performance. An AUC $>50\%$ indicates a better than chance performance, and an AUC closer to $100\%$ indicates the ideal model performance. We also compute the model specificity at $80\%$ sensitivity.

\section{Baseline System}
\subsection{Data preparation}
The audio data is pre-processed by normalizing the amplitude range to $\pm 1$. Subsequently, a simple sample level sound activity detection (SAD) is applied. This keeps any audio sample with absolute value greater than $0.01$ (and a margin of $\pm50~$msec around it) and discards the rest of the audio samples. Further, the initial and the final $20~$msec audio samples are also discarded to remove abrupt start and end burst due to device noise.


\subsection{Feature Extraction}
Here, $39$ dimensional mel-frequency cepstral coefficients (MFCC) \cite{davis_mfcc} and the delta and delta-delta coefficients are extracted with a window of size $1024$ samples and a hop of size $441$ samples. The \texttt{librosa} python library  \cite{brian_mcfee_2020_3955228} is used for the computation.
\begin{figure*}[t]
    \centering
    \input{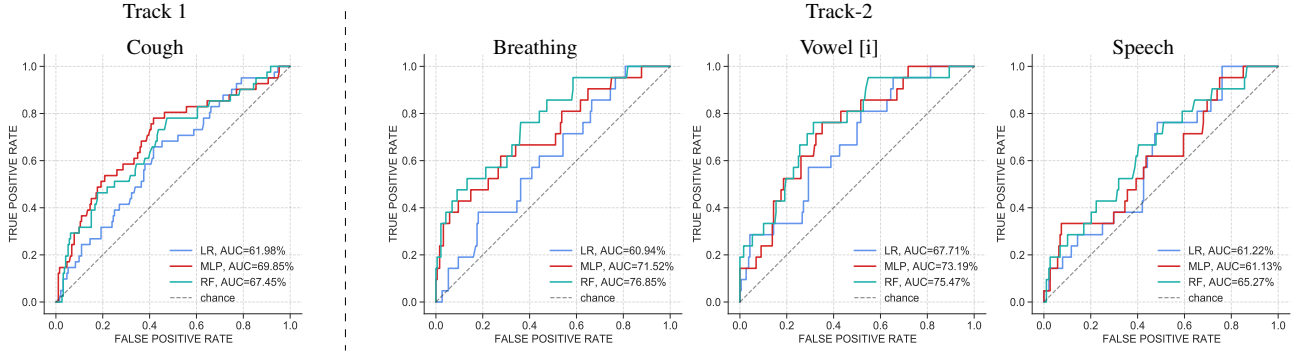}
    \caption{Illustration of baseline systems ROCs obtained on Track-1 and Track-2 test datasets.}
    \label{fig:test_track_1_perfm}
\end{figure*}

\subsection{Model Training}
Three different classifier models are trained for the two class classification tasks of COVID versus non-COVID detection.
The models are trained using the extracted features and a (class) balanced loss function, separately, for each of the five training data splits.
The implementation uses the \texttt{scikit-learn} python library \cite{scikit-learn}. The classifier models include the following.
\begin{itemize}
\item Logistic regression (LR): A logistic regression classifier trained with an added $\ell_2$ penalty, regularization strength of $0.01$ and \texttt{liblinear} optimizer is used. The maximum number of iterations is chosen as $25$.
\item Multi-layer perceptron (MLP): A single layer perceptron with $25$ hidden units, $\tanh()$ activation, and $\ell_2$ regularization penalty with a weight of $0.001$ is used. The loss function is optimised using the Adam optimizer with an initial learning rate of $0.001$. To implement balanced loss for MLP, the COVID class samples are randomly oversampled in appropriate proportion to match the count of non-COVID class samples.    
\item Random Forest (RF): The random forest classifier is trained with $50$ trees in the forest and \textit{gini} impurity criterion to measure the split quality.
\end{itemize}
 
\subsection{Model Inference and Decisions}
To obtain a classification score for an audio file: $(i)$ a pre-processing with amplitude normalization and SAD is done, $(ii)$ frame-level MFCC features are extracted, $(iii)$ frame-level probability scores are computed using the trained model, and $(iv)$ all the frame scores are averaged to obtain a single COVID probability score for the audio file.

\subsection{Results}
Table~\ref{tab:baseline_aucs} depicts the AUCs obtained on the validation folds. For each fold (shown in Fig.~\ref{fig:baseline_tracks_metadata}), the classifier is trained using the training data and evaluated on the validation data. The average validation AUC denotes the average over the AUCs for the five folds. For Track-1, RF gave the best average AUC, equating to $70.69\%$, and this was followed by MLP (at $68.80\%$) and LR (at $66.95\%$). For Track-2, RF gave superior performance on breathing sound ($75.17\%$~AUC). All models performed similar for vowel sound with an AUC close to $70\%$. The MLP gave a superior performance for speech sound, with $73.57\%$~AUC.

\begin{table}[]
\centering
\begin{tabular}{cllc}
\hline
\textbf{Track} & \textbf{Sound} & \multicolumn{1}{c}{\textbf{Model}} & \begin{tabular}[c]{@{}c@{}}\textbf{Avg.Val} \\ (Std. Err.)\end{tabular} \\ \hline
\textbf{} & \textbf{} & LR & 66.95 (±1.74) \\
1 & Cough & MLP & 68.54 (±1.65) \\
 &  & RF & 70.69 (±1.39) \\ \hline
\textbf{} & \textbf{} & LR & 60.95 (±2.17) \\
 & Breathing & MLP & 72.47 (±1.96) \\
 &  & RF & 75.17 (±1.23) \\ \cline{2-4} 
 & \textbf{} & LR & 71.48 (±0.55) \\
2 & Vowel {[}i{]} & MLP & 70.39 (±1.84) \\
 &  & RF & 69.73 (±1.93) \\ \cline{2-4} 
 & \textbf{} & LR & 68.93 (±1.09) \\
 & Speech & MLP & 73.57 (±0.71) \\
 &  & RF & 69.61 (±1.56) \\ \hline
\end{tabular}
\caption{The baseline system performance on the validation folds.}
\label{tab:baseline_aucs}
\end{table}

\noindent 
For evaluation on the test dataset, the COVID probability score for each file was computed by taking the average over the score outputs from the five validation fold models. The Track-1 blind test dataset release contains $233$ ($41$ COVID) cough audio files  for classification into COVID/non-COVID. For Track-1, the LR, MLP, and RF gave $61.98\%$, $69.85\%$, and $67.45\%$ AUCs, respectively. The corresponding ROCs are shown in Fig.~\ref{fig:test_track_1_perfm}.

The Track-2 test dataset release contains $209$ ($21$ COVID) audio files for each of the three sound categories. Here, the RF model gave a better performance than other models in all the three sound categories. Its performance was best for breathing ($76.85\%$ AUC) and worst for speech ($65.27\%$ AUC).

\section{Conclusion}
\label{conclusion}
The uniqueness of the dataset makes the DiCOVA challenge a first-of-its kind in the INTERSPEECH conference. The practical and timely relevance of the task encourages a focused effort from researchers across the globe, and from diverse fields such as respiratory sciences, speech and audio processing, and machine learning.
Along with the dataset, we also provide the baseline system software to all the participants. We expect this will serve as an example data processing pipeline for the participants. Further, participants are encouraged to explore different kinds of features and models of their own choice to obtain significantly better performance compared to the baseline system.

\section{Acknowledgement}
We thank Anand Mohan for his enormous help in web design and data collection efforts.
\bibliographystyle{IEEEtran}

\bibliography{mybib}

\end{document}